\documentclass[runningheads]{llncs}
\usepackage[T1]{fontenc}
\usepackage{cite}
\usepackage{amsmath,amssymb,amsfonts}
\usepackage{graphicx} \graphicspath{ {figures/} }
\usepackage{microtype}
\usepackage{makecell}
\usepackage{multirow}
\usepackage{comment}
\usepackage{hhline}
\usepackage{arydshln}
\usepackage{subfig}
\usepackage{balance}
\usepackage{booktabs}
\usepackage{tikz}

\usepackage[hidelinks]{hyperref}

\newcommand{\ack}{\subsubsection{\ackname}}

\begin{document}
\title{Cross-Slice Attention and Evidential Critical Loss for Uncertainty-Aware Prostate Cancer Detection}
\titlerunning{GLCSA and EC Loss for Uncertainty-Aware PCa Detection}
%
\author{Alex Ling Yu Hung\inst{1,2}\orcidID{0000-0002-2664-7430}, Haoxin Zheng\inst{1,2}\orcidID{0000-0002-8081-1940}, Kai Zhao\inst{1}\orcidID{0000-0002-2496-0829}, Kaifeng Pang\inst{1,3}\orcidID{0009-0001-7613-6559},\\ Demetri Terzopoulos\inst{2,4}, Kyunghyun Sung\inst{1}\orcidID{0000-0003-4175-5322}}
%
\authorrunning{A.~Hung et al.}
%
\institute{Department of Radiological Science, UCLA \and 
Computer Science Department, UCLA \and
Electrical and Computer Engineering Department, UCLA \and 
VoxelCloud, Inc.\\
\email{alexhung96@ucla.edu}}
\maketitle              
\begin{abstract} 
Current deep learning-based models typically analyze medical images in either 2D or 3D albeit disregarding volumetric information or suffering sub-optimal performance due to the anisotropic resolution of MR data. 
Furthermore, providing an accurate uncertainty estimation is beneficial to clinicians, as it indicates how confident a model is about its prediction. 
We propose a novel 2.5D cross-slice attention model that utilizes both global and local information, along with an evidential critical loss, to perform evidential deep learning for the detection in MR images of prostate cancer, one of the most common cancers and a leading cause of cancer-related death in men. 
We perform extensive experiments with our model on two different datasets and achieve state-of-the-art performance in prostate cancer detection along with improved epistemic uncertainty estimation. 
The implementation of the model is available at \url{https://github.com/aL3x-O-o-Hung/GLCSA_ECLoss}.
\end{abstract}

\section{Introduction}

Prostate cancer (PCa) is one of the most common cancers and a leading cause of cancer-related death in men~\cite{rawla2019epidemiology}. 
Multi-parametric magnetic resonance imaging (mpMRI) is commonly used for PCa diagnosis before conducting a prostate biopsy~\cite{appayya2018national}. 
General 2D approaches such as UNet~\cite{ronneberger2015u} and Swin-UNet~\cite{cao2022swin} along with PCa detection specific approaches such as FocalNet~\cite{cao2019joint} and ProstAttention-Net~\cite{duran2022prostattention} perform decently, but they fail to incorporate volumetric information. 
Although 3D methods such as 3D UNet~\cite{cciccek20163d} and VoxResNet~\cite{Chen2018VoxResNet} are able to achieve satisfactory results in MRI PCa detection, 3D models inherently struggle against anisotropic data~\cite{jia20193d,isensee2018nnu}. 
In prostate mpMRI, the through-plane resolution (3--6\,mm) is typically much poorer than the in-plane resolution (0.3--1.0\,mm)~\cite{turkbey2019prostate}, which poses challenges that undermine the ability of 3D models to perform accurate analysis. 
2.5D methods, including CAT-Net~\cite{hung2022cat}, CSAM~\cite{hung2024csam}, and AFTER-UNet~\cite{yan2022after}, among others, were proposed to address these challenges. 
They primarily analyze images in 2D, while incorporating volumetric information through an attention mechanism, 3D convolution, concatenation, etc. 
This aligns with how radiologists analyze images to make clinical decisions, inasmuch as they mainly focus on an image of interest while also attending to and considering other slices in the volume. 
Nevertheless, these methods under-utilize the local information and have many hyperparameters to tune and/or parameters to train. 

Despite the success of current deep learning-based approaches in MRI PCa detection in terms of accuracy and detection rate, there remains room for improvement.
Measuring the uncertainty of predictions is crucial for enhancing the robustness of the models, potentially aiding radiologists in making more informed diagnoses and decisions, because it offers insights into how confident the network is about its per-pixel predictions. 
In particular, epistemic uncertainty is more important as it depicts model uncertainty~\cite{kendall2017uncertainties}. 
Uncertainty in segmentation has been discussed in medical imaging~\cite{tang2022unified,carannante2021trustworthy}, but studies have focused on larger organs rather than much smaller cancer lesions. 
Moreover, most current research has employed techniques such as Monte Carlo (MC) dropout, ensemble, or an auxiliary branch to measure the epistemic uncertainty while relying on \textit{softmax} for class probabilities even though this is known to inflate the probability of the predicted class due to the exponentiation of the outputs~\cite{sensoy2018evidential}. 
  
We address this problem from two perspectives. 
First, we aim to improve cross-slice attention by considering both global and local information. 
Second, we perform evidential deep learning (EDL) on PCa detection for improved epistemic uncertainty estimation. 
The main contributions of the work are as follows:
\begin{enumerate}
\item We devise a novel cross-slice attention mechanism, named Global-Local Cross-Slice Attention (GLCSA), leveraging both local and global information.
\item We introduce a new evidential critical (EC) loss for EDL for PCa detection, which to our knowledge is the first EDL method applied to the detection or segmentation of small objects. 
\item We achieve state-of-the-art (SOTA) results across two different PCa datasets. 
\item We show that our method for estimating epistemic uncertainty outperforms competing approaches, which can lead to improved detection results.
\end{enumerate}

\section{Preliminaries}

\subsection{2.5D Volumetric Image Analysis}

We denote an input image volume $x_0\in\mathbb{R}^{l\times c_0\times h_0\times w_0}$, where $l$ is the total number of slices in the volume, $c_0$ is the number of channels, indicating the imaging modalities or types of imaging that are available, and $h_0$ and $w_0$ are the height and width of each slice. 
The pure 2D encoder $E$ and decoder $D$ in a standard encoder-decoder architecture for image segmentation are accompanied by three functions: $f_{\text{pre}}(\cdot)$ applied to the original volume, $f_{\text{mid}}(\cdot)$ applied to the feature maps between the encoder and decoder, and $f_{\text{post}}(\cdot)$ applied to the output of the decoder. 
2.5D volumetric image segmentation is defined as follows: 
\begin{quote}
For any $E$ and $D$, given the segmentation output
\begin{equation}
    y=f_{\text{post}}(D(f_{\text{mid}}(E(f_{\text{pre}}(x_0))))),
    \label{eq:general}
\end{equation}
when at least one of $f_{\text{pre}}(\cdot)$, $f_{\text{mid}}(\cdot)$, and $f_{\text{post}}(\cdot)$ involves operations between different slices, the segmentation model is 2.5D~\cite{hung2024csam}. 
\end{quote}

Conventional 2.5D models stack nearby slices as input to segment the slice of interest~\cite{han2021liver}. 
In these methods $f_\text{pre}(\cdot)$ stacks nearby slices, while $f_\text{mid}(\cdot)$ and $f_\text{post}(\cdot)$ are identity functions.
In contrast, SAUNet~\cite{zhang2020sau} performs attention across different slices in $f_\text{post}(\cdot)$, with $f_\text{pre}(\cdot)$ and $f_\text{mid}(\cdot)$ being identity functions.
More recent works, such as CAT-Net~\cite{hung2022cat}, AFTER-UNet~\cite{yan2022after}, CSAM~\cite{hung2024csam}, etc., conduct cross-slice attention between the encoder and decoder, where $f_\text{mid}(\cdot)$ is the cross-slice attention module, while $f_\text{pre}(\cdot)$ and $f_\text{post}(\cdot)$ are identity functions.

\subsection{Evidential Deep Learning}

Sensoy et al.~\cite{sensoy2018evidential} proposed EDL based on the Dempster–Shafer Theory (DST) of Evidence~\cite{shafer1976mathematical,shafer1992dempster} to explicitly model the epistemic uncertainty of each point prediction by assigning a set of belief masses $b_k \geq 0$ for any $k \in [1,K]$ representing a set of exclusive possible states; i.e., the possible class labels at a given pixel. 
The belief masses, each of which is associated with the Dirichlet distribution, along with the overall uncertainty mass $u\geq0$ (the ``I don't know'' class) 
should sum to 1:
\begin{equation}
u+\sum_{k=1}^Kb_k=1.
\end{equation}
Belief $b_k$ for singleton $k$ and uncertainty $u$ may be calculated as
\begin{equation}
b_k=\frac{e_k}{S}, \qquad u=\frac{K}{S}, \qquad S=K+\sum_{k=1}^Ke_k,
\label{eq:belief and uncertainty}
\end{equation}
where $e_k$ is the evidence, which quantifies the level of support gathered from the data for classifying the sample into class $k$. 
Equation~\eqref{eq:belief and uncertainty} demonstrates that increased evidence leads to decreased uncertainty, and 
in the absence of any evidence, $u=1$, indicating complete uncertainty.  
This is associated with the Dirichlet distribution by $\alpha_k=e_k+1$, where $\alpha_k$ is the Dirichlet concentration parameter for class $k$. 
The Dirichlet distribution is the conjugate prior of the categorical distribution, so the probability assignment of the categorical distribution over all possible classes can be sampled according to 
\begin{equation}
\Vec{p}\sim\text{Dir}(\Vec{p}|\Vec{\alpha}), \qquad \Vec{\hat{y}}\sim\text{Cat}(\Vec{y}|\Vec{p}),
\end{equation}
where $\text{Dir}(\Vec{p}|\Vec{\alpha})$ is the Dirichlet distribution over probabilities $\Vec{p}$ given concentration parameters $\Vec{\alpha}=(\alpha_1,\alpha_2, \dots,\alpha_K)$ and $\text{Cat}(\Vec{y}|\Vec{p})$ is the categorical distribution over classes $\Vec{y}$ in one-hot encoding given the probabilities for each class $\Vec{p}$. 
The expected probability of singleton $k$ is the mean of the corresponding Dirichlet distribution: 
\begin{equation}
\hat{p}_k=\frac{\alpha_k}{S}.
\end{equation}

EDL networks replace the conventional \textit{softmax} function in the final layer with an alternative activation function whose output is taken as the evidence vector $\Vec{e}=(e_1, e_2, \dots, e_K)$, which is then used to calculate the concentration parameters $\Vec{\alpha}$ that parameterize the Dirichlet distribution. 
Thus, standard deep learning directly predicts the class probability distribution, while EDL predicts the density of the class probability in the categorical distribution by modeling a Dirichlet distribution. 
As a result, EDL models the second-order probabilities, giving us more reasonable class probabilities and uncertainty estimation. 

\section{Global-Local Cross-Slice Attention}

\begin{figure}[t] \centering
\subfloat[GLCSA in nnUNet\label{fig:glcsa-nnunet}]
{\begin{minipage}{.29\textwidth}
\scalebox{0.29}{
    \begin{tikzpicture}
\node [circle,draw=black,fill=black!30,inner sep=0pt,minimum size=1cm] (x00) at (-7.6,-2) {$X_{0,0}$};
\node [circle,draw=black,fill=black!30,inner sep=0pt,minimum size=1cm] (x10) at (-6.7,-3.5) {$X_{1,0}$};
\node [circle,draw=black,fill=black!30,inner sep=0pt,minimum size=1cm] (x20) at (-5.8,-5) {$X_{2,0}$};
\node [circle,draw=black,fill=black!30,inner sep=0pt,minimum size=1cm] (x30) at (-4.9,-6.5) {$X_{3,0}$};
\node [circle,draw=black,fill=black!30,inner sep=0pt,minimum size=1cm] (x40) at (-4,-8) {$X_{4,0}$};

\node [rectangle,draw=black,fill=white,inner sep=0pt,minimum size=1cm] (isim0) at (-3,-2) {$GLCSA_0$};
\node [rectangle,draw=black,fill=white,inner sep=0pt,minimum size=1cm] (isim1) at (-3,-3.5) {$GLCSA_1$};
\node [rectangle,draw=black,fill=white,inner sep=0pt,minimum size=1cm] (isim2) at (-3,-5) {$GLCSA_2$};
\node [rectangle,draw=black,fill=white,inner sep=0pt,minimum size=1cm] (isim3) at (-3,-6.5) {$GLCSA_3$};
\node [rectangle,draw=black,fill=white,inner sep=0pt,minimum size=1cm] (isim4) at (-2,-8) {$GLCSA_4$};

\node [circle,draw=black,fill=white,inner sep=0pt,minimum size=1cm] (x31) at (-1.1,-6.5) {$X_{3,1}$};

\node [circle,draw=black,fill=white,inner sep=0pt,minimum size=1cm] (x21) at (-0.2,-5) {$X_{2,1}$};

\node [circle,draw=black,fill=white,inner sep=0pt,minimum size=1cm] (x11) at (0.7,-3.5) {$X_{1,1}$};

\node [circle,draw=black,fill=white,inner sep=0pt,minimum size=1cm] (x01) at (1.6,-2) {$X_{0,1}$};

\path [draw,->] (x00)edge[] (isim0);
\path [draw,->] (isim0)edge[densely dotted] (x01);

\path [draw,->] (x10)edge[] (isim1);
\path [draw,->] (isim1)edge[densely dotted] (x11);

\path [draw,->] (x20)edge[] (isim2);
\path [draw,->] (isim2)edge[densely dotted] (x21);

\path [draw,->] (x30)edge[] (isim3);
\path [draw,->] (isim3)edge[densely dotted] (x31);

\path [draw,->] (x40)edge[] (isim4);

\path [draw,->] (isim4)edge[] (x31);

\path [draw,->] (x00)edge[] (x10);
\path [draw,->] (x10)edge[] (x20);
\path [draw,->] (x20)edge[] (x30);
\path [draw,->] (x30)edge[] (x40);

\path [draw,->] (x21)edge[] (x11);
\path [draw,->] (x31)edge[] (x21);
\path [draw,->] (x11)edge[] (x01);
\path [draw,->] (x01)edge[] (3,-2);
\coordinate (temp) at (-9,-2);
\draw[->] (temp) -- (x00);
\end{tikzpicture}
}
\end{minipage}}
\hfill
\subfloat[GLCSA Pipeline\label{fig:glcsa-pipeline}]
{\begin{minipage}{.69\textwidth}
\centering
    \includegraphics[width=\textwidth]{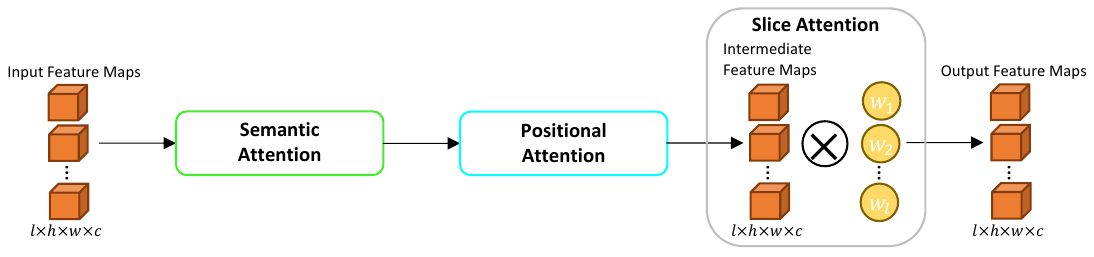}
\end{minipage}}\\
\subfloat[Semantic Attention\label{fig:semantic}]{\includegraphics[width=0.49\textwidth]{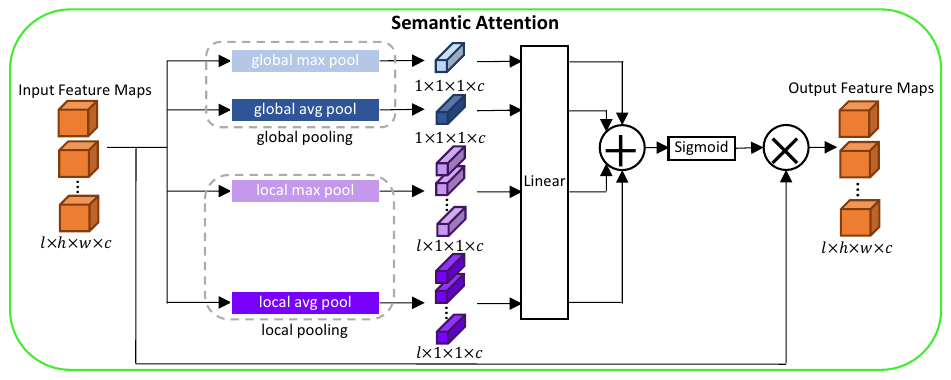}} 
\hfill
\subfloat[Positional Attention\label{fig:positional}]{\includegraphics[width=0.49\textwidth]{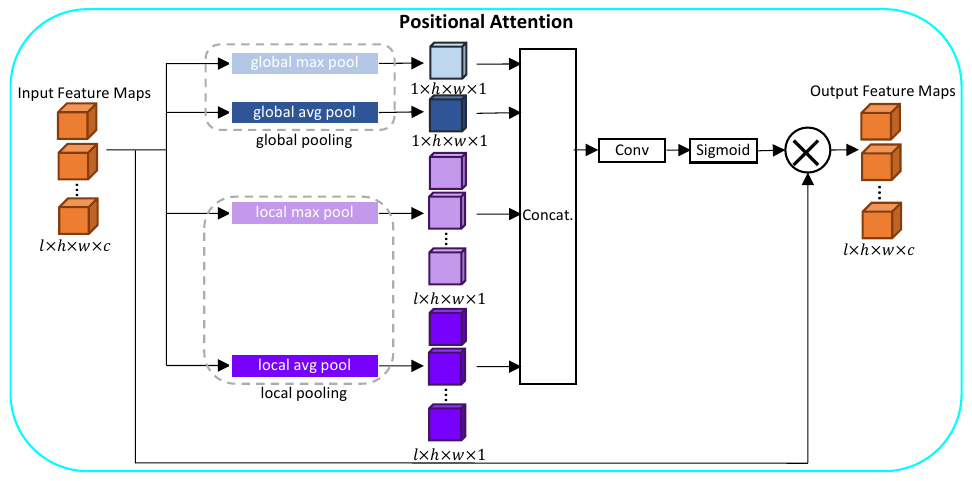}}
\caption{GLCSA implementation.}
\label{fig:implementation}
\end{figure}
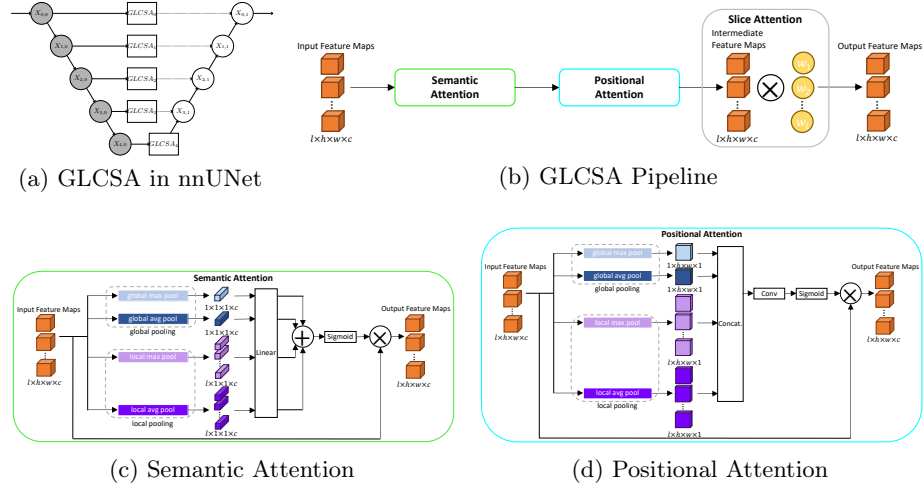

GLCSA has three parts: semantic, positional, and slice attention, organized sequentially. 
It can be inserted into any UNet-like architecture at the skip connection. 
2.5D GLCSA processes the feature maps generated by the encoder and forwards its outputs to the decoder; i.e., GLCSA is $f_{\text{mid}}(\cdot)$ in \eqref{eq:general}. 
\autoref{fig:glcsa-nnunet} illustrates the implementation of GLCSA in UNet-like architectures and \autoref{fig:glcsa-pipeline} illustrates the GLCSA pipeline.

\subsection{Semantic Attention}
As shown in \autoref{fig:semantic}, semantic attention processes input feature maps to re-weight them based on the relevance of semantic information across all the 2D feature maps of the volume.
The input feature maps have dimensions $l\times h\times w\times c$, where $l$ is the number of slices in the volume, $h$ and $w$ are the height and width, and $c$ is the number of channels. 
To perform semantic attention, the feature maps are first squeezed in the $l$, $h$, and $w$ dimensions for global max and average pooling, and in the $h$ and $w$ dimensions for local max and average pooling. 
The outputs of the pooling layers are then fed into a linear layer and subsequently added together before going into the sigmoid function to produce the attention maps.
 
\subsection{Positional Attention}
Positional attention, illustrated in \autoref{fig:positional}, takes as input the output of semantic attention and adjusts the feature maps according to the significance of positions within the feature maps of each slice. 
The input features maps are first squeezed in the $l$ and $c$ dimensions for global max and average pooling, and in only the $c$ dimension for local max and average pooling. 
They are then concatenated in the $c$ dimension before passing a learnable convolutional kernel, whose output passes through a sigmoid function to produce the attention map that re-weights the input feature maps. 

\subsection{Slice Attention}
Slice attention, illustrated in \autoref{fig:glcsa-pipeline}, weights each slice based on the relative position of the slice. 
For instance, slices at the top and bottom extremities of the volume, should receive lower weights as they are less likely to intersect the prostate, whereas central slices should be assigned higher weights are they are more likely to span it. 
This attention mechanism employs learnable weights $w_i$, where $1\leq i\leq l$, each of which multiplies
the feature maps $F_i\in\mathbb{R}^{h\times w\times c}$ output from the positional attention for slice $i$.   
The re-weighted feature maps are then fed into the decoder. 

\section{Evidential Critical Loss}

EDL effectively quantifies epistemic uncertainty without extra computational cost; 
however, conventional EDL optimizes the Bayes risk of either cross-entropy loss or sum of squared loss~\cite{sensoy2018evidential}.
These training objectives fail to account for the disproportionately small size of PCa lesions relative to the imaging volume.
The number of background pixels outnumbers the lesion pixels by a factor of 1K to 10K in patients with suspicious lesions, making accurate PCa detection challenging and leading to potential overfitting to the background.
Therefore, we introduce a new training objective, the EC loss, for evidential deep learning on highly imbalanced datasets. 
This loss function prioritizes critical classes, i.e., classes that are hard to classify, by reducing the loss impact from easier examples. 

For a pixel $i$, we let $\Vec{y}_i$ and $\Vec{p}_i$ be the one-hot encoding vectors for the ground truth and prediction, respectively, and denote the concentration parameters as $\Vec{\alpha}_i$. 
The conventional evidential loss can be written as
\begin{equation}
    L_{\text{ev},i}=
    \int\left\|\Vec{y}_i-\Vec{p}_i\right\|_2^2
    \frac{\prod_{j=1}^{K}p_{ij}^{\alpha_{ij}-1}}{B(\Vec{\alpha}_i)}d\Vec{p}_i=
    \sum_{j=1}^{K}\left((y_{ij}-p_{ij})^2+\frac{p_{ij}(1-p_{ij})}{S_i+1}\right),
\end{equation}
where $K$ is the number of classes, $B(\cdot)$ is the Beta function, and $S_i=\sum_{j=1}^K\alpha_{ij}$.
However, this would result in the model overfitting the easy classes. 
Therefore, a modulation factor $(1-\lambda_i)^\gamma$ and a weighting factor $\beta_i$ are introduced to prioritize the critical classes \cite{lin2017focal,cao2019joint}; i.e., emphasizing the loss at true lesion pixels: 
\begin{equation}
    L_{\text{EC},i}=(1-\lambda_i)^\gamma\beta_iL_{\text{ev},i},
\end{equation}
where $\lambda_i=\exp(\sum_{j=1}^Ky_{ij}\log p_{ij})$ is the likelihood of the ground truths given the model predictions and $\gamma$ controls the modulation intensity. 
When the likelihood $\lambda_i$ is high, the pixel is well-classified and the EC loss is down-weighted close to 0, whereas a low likelihood $\lambda_i$ suggests that the pixel is incorrectly classified,
and the EC loss is less affected by the modulation. 
 Thus, the EC loss targets critical pixels; i.e., those prone to incorrect classification. 
The modulation parameter $\gamma$ controls the rate at which the non-critical pixels are down-weighted, with larger values reducing the effect of non-critical pixels on the loss.  
Weighting factor $\beta_i$ controls which pixels the loss prioritizes, with higher values assigned to lesion pixels to increase the penalty for missed detection. 

False evidence may arise during training, but the EC loss alone cannot eliminate it.
Ideally, the total evidence should shrink to 0 if the model is not able to correctly classify the specific pixel.
Therefore, an additional regularizer is needed, which incorporates a Kullback-Leibler (KL) divergence into the training objective, which penalizes evidence that does not contribute to the data fit. 
Finally, the total loss is formulated as
\begin{equation}
    L=\sum_{i=1}^NL_{\text{EC},i}+s\sum_{i=1}^NKL\left(\text{Dir}(\Vec{p}_i|\Vec{\alpha}_i)||\text{Dir}(\Vec{p}_i|\Vec{1})\right),
\end{equation}
where $s$ is the annealing coefficient during training, $KL(\cdot||\cdot)$ is the KL divergence between two distributions, and $\Vec{1}$ is a vector of ones. 

\section{Experiments}

\subsection{Datasets}
We used two datasets, including our internal dataset (\textit{Internal}) and the PI-CAI challenge~\cite{PICAI_BIAS} dataset (\textit{PICAI}). 
\textit{Internal} comprises 652 patients with T2-weighted imaging (T2WI), apparent diffusion coefficient (ADC), high $b$-value (HBV) ($b=1400$) diffusion-weighted imaging (DWI) MRI, along with manual zonal segmentation annotations. 
T2WI, ADC, HBV, and zonal segmentation are stacked as the model input.
\textit{PICAI} comprises 1,293 patients who underwent T2WI, ADC, HBV ($b>1000$) DWI MRI, annotated by clinical experts. 
Only T2WI, ADC, and HBV are stacked as the model input. 

\subsection{Implementation Details}
We implemented our method on nnUNet~\cite{isensee2018nnu} (\autoref{fig:implementation}).
We use the segmentation model to perform PCa detection. 
The local maxima in the output probability map are considered the cancer detection points. 
The true positive PCa detection point is defined when it is within 5\,mm of any PCa ground truth so as to account for a potential mismatch between the real lesion and the label due to labeling errors.
For the EC loss, $\beta_i=30$ if pixel $i$ is a lesion pixel and $\beta_i=1$ if it is a background pixel. 
The value for $\beta_i$ is from the weighting in focal loss provided in AtPCa-Net~\cite{zheng2024atpca}.
On \textit{Internal} $\gamma=2.5$, while on \textit{PICAI} $\gamma=4$. 
We utilized focal loss~\cite{lin2017focal} for all the other experiments, and we set the focusing parameter to 2 for \textit{Internal} and to 3 for \textit{PICAI}. 
We used the Adam optimizer with a learning rate of 0.0001 and weight decay regularization set to $1\times10^{-5}$. 
For all experiments, we used 5-fold cross-validation. 

\begin{figure}[t] \centering
\subfloat[FROC on \textit{Internal}]
{\includegraphics[width=0.49\textwidth]{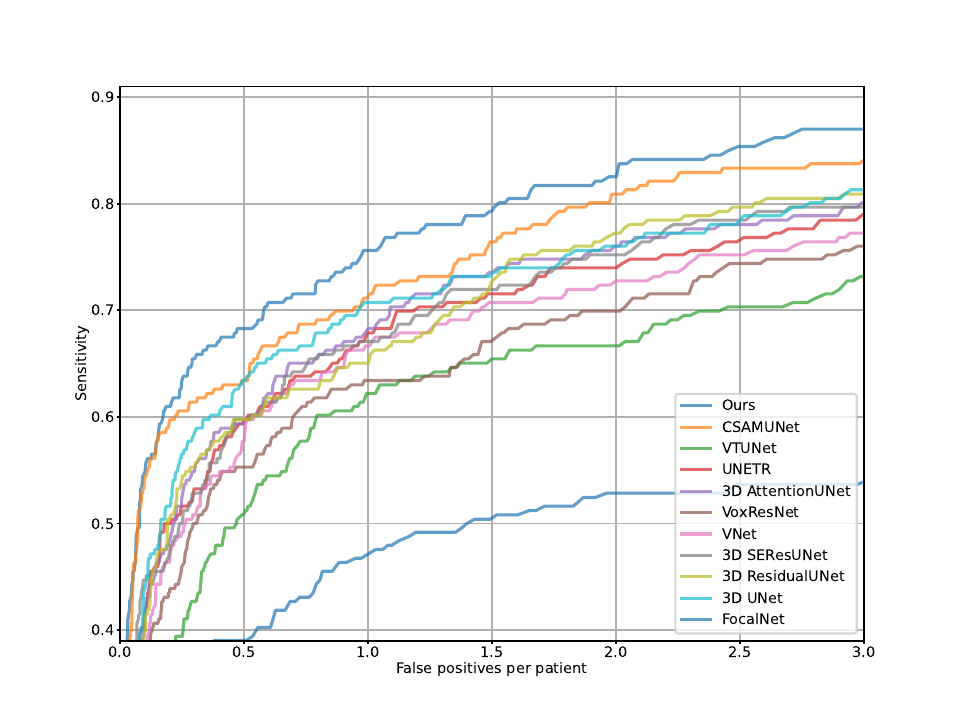}}
\hfill
\subfloat[FROC on \textit{PICAI}]
{\includegraphics[width=0.49\textwidth]{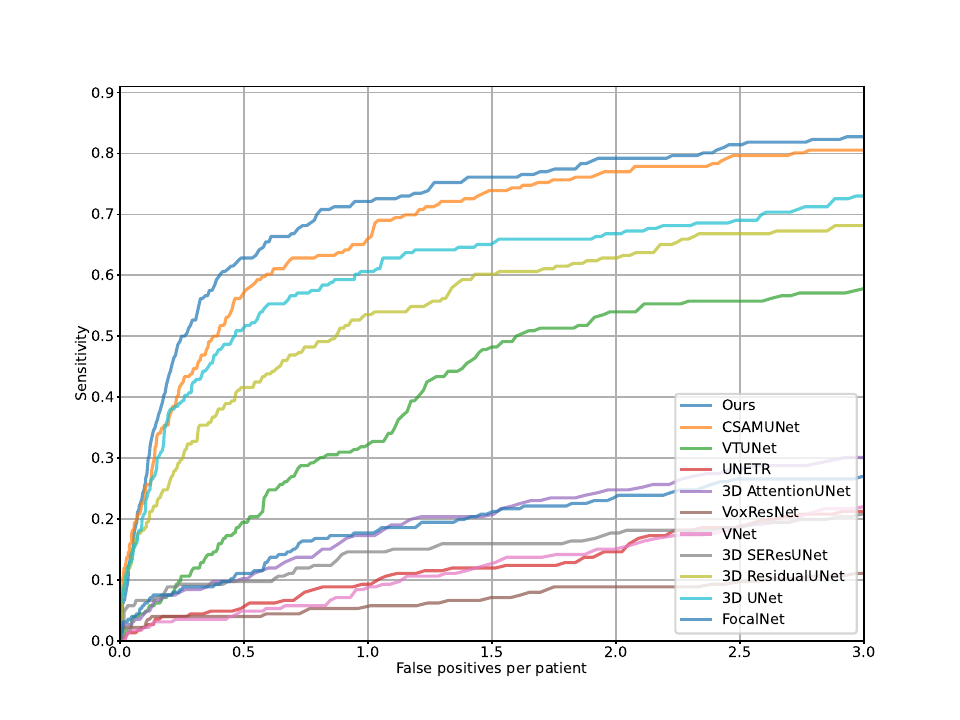}}
\caption{Quantitative results.}
\label{fig:quantitative}
\end{figure}

\subsection{Quantitative Results}
The Free-response Receiver Operating Characteristic (FROC) curves in \autoref{fig:quantitative} show that for both datasets our method outperforms all the competing methods; i.e., FocalNet~\cite{cao2019joint}, 3D UNet~\cite{cciccek20163d}, 3D ResidualUNet~\cite{Bhalerao2020ResUNet}, 3D SEResUNet~\cite{Hu2018SENet}, 3D AttentionUNet~\cite{oktay2018attention}, VNet~\cite{milletari2016v}, VoxResNet~\cite{Chen2018VoxResNet}, UNETR~\cite{hatamizadeh2022unetr}, and VTUNet~\cite{peiris2022robust}. 
The absence of zonal segmentation in the inputs in \textit{PICAI} hampers many competing methods~\cite{hosseinzadeh2022deep}, whereas
our method consistently delivers the best results regardless of the availability of zonal segmentation. Our supplementary materials include qualitative results.  

\subsection{Ablation Study}
\autoref{table:ablation} reports the results of our ablation study on \textit{Internal}, comparing PCa detection sensitivities at different false positives (FPs)/patient. GCSA indicates global cross-slice attention; i.e., GLCSA without the local poolings. 
By introducing in turn GCSA, EC loss, and local poolings, the study confirmed that each component plays a role in improving performance. 

\begin{table*}[t] \setlength{\tabcolsep}{5pt}
\caption{Ablation study on \textit{Internal}.}
\centering
\begin{tabular}{ lc ccccc }
\toprule
FPs/Patient&
 0.5&1&1.5&2& 2.5&3\\
\midrule
nnUNet&0.659&0.715&0.740&0.768&0.789&0.801\\
nnUNet+GCSA&0.654&0.724&0.772&0.813&0.833&0.841\\
nnUNet+GCSA+EC loss&0.671&\bfseries0.760&\bfseries0.805&0.813&0.833&0.850\\
nnUNet+GLCSA+EC loss&\bfseries0.687&\bfseries0.760&0.801&\bfseries0.841&\bfseries0.854&\bfseries0.870\\
\bottomrule
\end{tabular}
\label{table:ablation}
\end{table*}

\begin{table*}[t] \setlength{\tabcolsep}{5pt}
\caption{Performances of different uncertainty methods on \textit{PICAI}.}
\centering
\begin{tabular}{ lc ccccc }
\toprule
FPs/Patient&
 0.5&1&1.5&2& 2.5&3\\
\midrule
MC Dropout&0.540&0.677&0.717&0.743&0.752&0.779\\
EDL&\bfseries0.642&\bfseries0.726&\bfseries0.761&\bfseries0.792&\bfseries0.819&\bfseries0.827\\
\bottomrule
\end{tabular}
\label{table:uncertainty}
\end{table*}

\begin{figure}[t]
\subfloat[\label{fig:uncertainty-plots}]
{\includegraphics[width=0.49\textwidth]{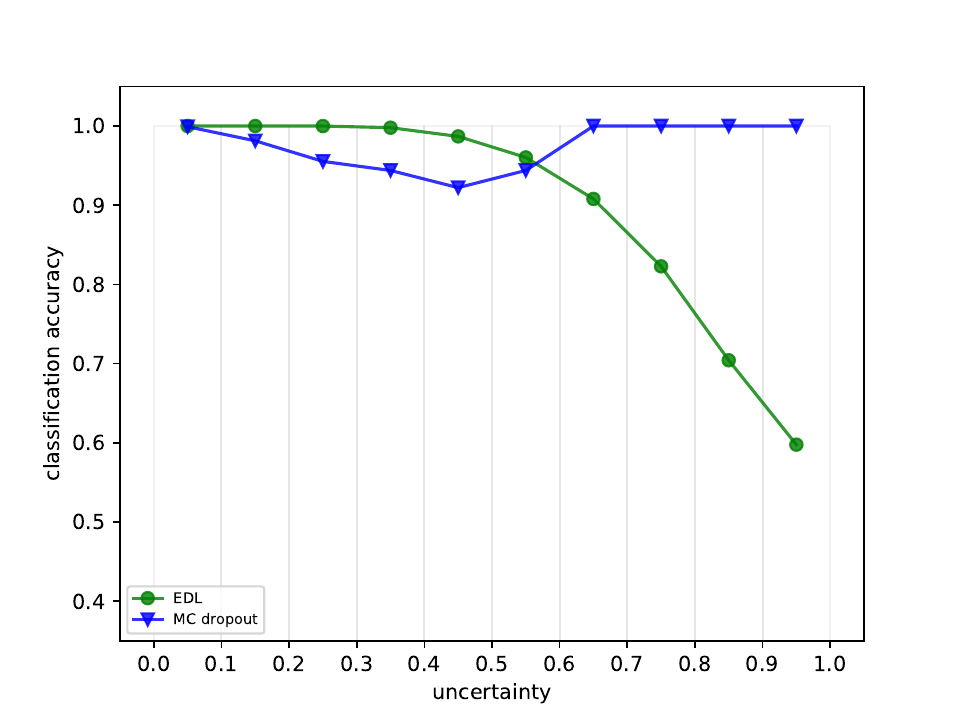}}
\hfill
\subfloat[\label{fig:froc}]
{\includegraphics[width=0.49\textwidth]{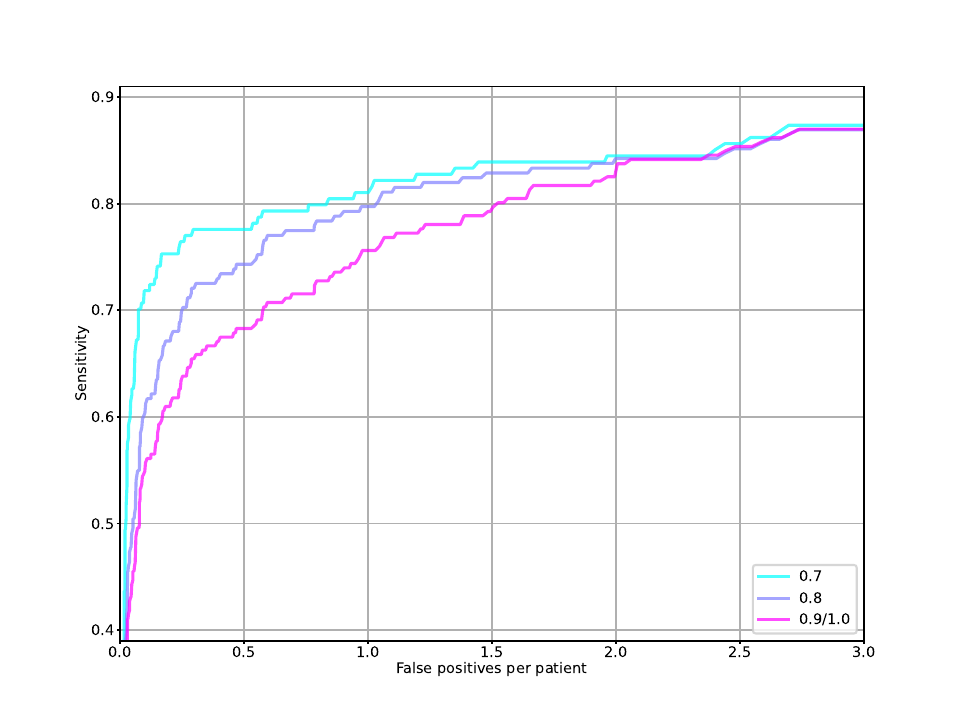}}
\caption{(a) Comparison of EDL and MC dropout on uncertainty calibration, and (b) FROC curves for different uncertainty thresholds.}
\label{fig:uncertainty}
\end{figure}

\subsection{Uncertainty Estimation}
Applying our proposed architecture on \textit{PICAI}, \autoref{table:uncertainty} shows that our EDL-based method outperforms the MC dropout-based methods. The supplementary materials include visual results of different uncertainty estimations on \textit{Internal}. 

Furthermore, we evaluated the effectiveness of our EDL-based method. 
The pixels were first grouped into 10 bins based on their uncertainty values, where each bin has a 0.1 interval in uncertainty. 
Then, the classification accuracy for each bin was calculated. 
Ideally, the accuracy should drop as the uncertainty increases. 
The results on \textit{Internal} are shown in \autoref{fig:uncertainty-plots}, where EDL achieves better uncertainty estimation since the accuracy drops as uncertainty increases, whereas uncertainty by MC dropout does not provide information regarding the classification result. 
Our method not only has better PCa detection performance, but also better uncertainty estimation than MC dropout.

Additionally, we investigated how uncertainty estimation can further help with PCa detection. 
We set thresholds such that any pixel with a higher uncertainty value is disregarded during PCa detection. 
We tested thresholds 0.7, 0.8, 0.9, and 1.0, which is equivalent to no threshold. 
The plots in \autoref{fig:froc} show that when using a lower threshold---i.e., eliminating more pixels---the sensitivity increases when the FPs/patient is relatively low. 
However, when the FPs/patient increases, the difference in sensitivity for different thresholds decreases. 
Thus, our uncertainty estimation can successfully tell where the model may be wrong, and that information can further improve detection sensitivity. 

\section{Conclusions}

We have introduced GLCSA, a novel cross-slice attention mechanism that utilizes both global and local information, as well as a new EC loss that enables EDL in PCa detection. 
Experimenting with two different datasets, we showed that our proposed method outperforms other SOTA methods in terms of FROC under simple training settings with data augmentation. 
Our ablation study demonstrated that all key components contributed to improving performance. 
The uncertainty estimation inherent to our method is superior to its counterparts as it can improve the explainability and interpretability of the model while improving its performance when some pixels are filtered out by thresholding uncertainty values.

\ack
The research reported herein was funded in part by the National Institutes of Health under grants R01-CA248506 and R01-CA272702 and by the Integrated Diagnostics Program of the Departments of Radiological Sciences and Pathology in the UCLA David Geffen School of Medicine.

\bibliographystyle{splncs04}
\bibliography{ref}
\end{document}